\documentstyle[amssymb,prl,aps,preprint]{revtex}

\begin{document}
\title{Symmetry of anisotropic exchange interactions in semiconductor nanostructures}
\author{K.V.Kavokin}
\address{{\it School of Physics, University of Exeter, Stocker Road, Exeter EX4 4QL,}%
\\
UK, and {\it A. F. Ioffe Physico-Technical Institute, 194021}\\
Politechnicheskaya 26, \\
St. Petersburg, Russia}
\maketitle

\begin{abstract}
The symmetry of exchange interaction of charge carriers in semiconductor
nanostructures (quantum wells and quantum dots) is analysed. It is shown
that the exchange Hamiltonian of two particles belonging to the same energy
band can be universally expressed via pseudospin operators of the particles.
The relative strength of the anisotropic exchange interaction is shown to be
independent of the binding energy and the isotropic exchange constant.
\end{abstract}

\narrowtext

The reduced symmetry of semiconductor nanostructures suggests that the
exchange interaction of charge carriers in such structures is not
necessarily described by the isotropic (Heisenberg) spin Hamiltonian. In
particular, the exchange interaction of electrons and holes in quantum wells
and quantum dots is known to be extremely anisotropic, giving rise to a fine
structure of nanostructure excitons \cite{IvchPik}. It has been shown \cite
{Moriarty} that the exchange interaction of conduction-band electrons is
also anisotropic if the structure lacks inversion symmetry. The main term of
the anisotropic exchange Hamiltonian in this case has the
Dzyaloshinskii-Moriya form\cite{Dzialoshinskii}. The electron-electron
anisotropic exchange was subsequently widely discussed in relation to the
quantum computing problem \cite{Bonesteel,BurLoss,LidarWu}. It has been
recently detected experimentally via its contribution to the spin relaxation
of donor-bound electrons in GaAs \cite{Taus}, where it has been shown to put
the upper boundary for the electron spin lifetime at donor concentrations
around 10$^{16}$ cm$^{-3}$ . However, it remained so far unclear whether or
not the anisotropic spin Hamiltonian suggested in Ref.\cite{Moriarty} is
universal for all types of charge carriers, e.g. for two-dimensional holes.
The issue of the dependence of the anisotropic exchange constant on the
parameters of the localizing potential is also very sensitive, especially
for the discussion on feasibility of quantum computation with solid-state
spin systems \cite{BurLoss}. The constant was so far calculated using the
Heitler-London method \cite{Moriarty,BurLoss}, which is known to give
incorrect asymptotic expression for the isotropic exchange integral \cite
{GorPit,HerFlick}. Gor'kov and Krotkov \cite{GK}, using the median-plane
method\cite{GorPit}, have recently obtained a correct asymptotic formula for
the anisotropic exchange constant in a specific case of hydrogen-like
centers in zinc-blende semiconductors, different from that calculated
earlier by the Heitler-London method \cite{Moriarty}. However their approach
is not always applicable to coupled quantum dots, where the distance between
quantum dots can be comparable to the quantum-dot size.

The collection of unsolved problems and blanc spaces in the existing
knowledge on the anisotropic exchange in semiconductor structures, given
above, demonstrates the evident demand for a consistent theoretical analysis
of the issue, based on a general approach. In this paper, we consider
exchange interaction of two identical charge carriers localized in any
symmetric double-well potential in a two-dimensional semiconductor
structure. Using the pseudospin formalism allows to obtain a universal spin
Hamiltonian describing this class of systems.

Let us consider the exchange interaction of two identical charge carriers
(electrons or holes), localized in two centrosymmetric potential hollows
(further referred to as quantum dots, QDs) in a quasi-two-dimensional
semiconductor structure (quantum well, QW). The QDs may be, for example,
self-organized QDs \cite{Ledenets}; otherwise, they can be induced by
electrostatic potential of nanometer-sized gates \cite{GateQD} or impurity
centers \cite{Efros}. The distance between centers of the QDs will further
be denoted as $R_{12}$. In quasi-two-dimensional structures, the 4-fold
degeneration of the valence band, typical of cubic semiconductors, is
lifted. The states at extremum points of two-dimensional subbands in absence
of magnetic fields retain only the Kramers two-fold degeneration. Their wave
functions can be written as $\Psi ({\bf r})u_{\nu }({\bf r})$, where is an
envelope function, and $u_{\nu }({\bf r})$ is a Bloch amplitude, $\nu =\pm
1/2$. The Bloch amplitudes $u_{\nu }({\bf r})$ are transformed into each
other by the operator of time reversal \cite{Landau}:

\begin{equation}
u_{-1/2}({\bf r})=-i\hat{\sigma}_{y}u_{+1/2}({\bf r})  \label{1}
\end{equation}

This property allows to associate the Kramers index $\nu $ with an
eigenvalue of a projection of a pseudospin operator{\bf \ }${\bf j}$ ($j=1/2$%
) on some (generally, fictitious) axis. The choice of basis functions for $%
{\bf j}$ is not unambiguous. It is limited only by the condition given by
Eq.(1). In particular, for heavy holes with the projection of the angular
momentum on the structure axis Z, equal to $J_{z}=\pm 3/2$, it is convenient
to choose the functions as \cite{IvchPik}: $\left| j,+1/2\right\rangle
=\left| J_{z},-3/2\right\rangle $ and $\left| j,-1/2\right\rangle =\left|
J_{z},+3/2\right\rangle $. This choice allows to avoid phase multipliers
which would otherwise appear at wave functions in the pseudospin
representation. For conduction-band electrons, the pseudospin coincides with
the electron spin ${\bf s}$. Linear transformations of pseudospin wave
functions determined in the basis $\left\{ u_{+1/2},u_{-1/2}\right\} $ are
equivalent to rotations of usual spinor functions \cite{Landau}:

\begin{eqnarray}
u_{+1/2}^{\alpha \beta \gamma } &=&\exp (i\gamma /2)\left[ u_{+1/2}\exp
(i\alpha /2)\cos (\beta /2)+u_{-1/2}\exp (-i\alpha /2)\sin (\beta /2)\right]
\nonumber \\
u_{-1/2}^{\alpha \beta \gamma } &=&\exp (-i\gamma /2)\left[ -u_{+1/2}\exp
(i\alpha /2)\sin (\beta /2)+u_{-1/2}\exp (-i\alpha /2)\cos (\beta /2)\right]
\label{2}
\end{eqnarray}

where $\alpha $, $\beta $, and $\gamma $ are analogs of Euler angles.
Following the analogy, one can introduce the total pseudospin ${\bf I=j}_{1}+%
{\bf j}_{2}$. Indeed, the Gilbert space of two-pseudospin wave functions $%
A_{\mu \nu }u_{\mu }({\bf r}_{1})u_{\nu }({\bf r}_{2})$ breaks into two
subspaces invariant with respect to the simultaneous transformation of both
pseudospins along Eq.(2) with the same $\alpha $, $\beta $, and $\gamma $.
The basis functions of these subspaces, $\xi _{0}=(u_{+1/2}({\bf r}%
_{1})u_{-1/2}({\bf r}_{2})-u_{-1/2}({\bf r}_{1})u_{+1/2}({\bf r}_{2}))/\sqrt{%
2}$ and $\xi _{1M}$, equal to $(u_{+1/2}({\bf r}_{1})u_{-1/2}({\bf r}%
_{2})+u_{+1/2}({\bf r}_{1})u_{-1/2}({\bf r}_{2}))/\sqrt{2}$ ($M=0$) or $%
u_{\pm 1/2}({\bf r}_{1})u_{\pm 1/2}({\bf r}_{2})$ ($M=\pm 1$), are,
obviously, eigenfunctions of the operators $\hat{I}^{2}$ and $\hat{I}_{z}$.

The general form of the one-particle Hamiltonian of the two dimensional
charge carrier in the pseudospin representation is

\begin{equation}
\hat{H}_{1}=\frac{\hbar ^{2}}{2m}\hat{k}^{2}+V({\bf r})+{\bf h}({\bf k}%
)\cdot {\bf j}  \label{3}
\end{equation}

where the ''spin-orbit field'' ${\bf h}({\bf k})$ is a vector in the
pseudospin space \cite{OO}. ${\bf h}({\bf k})$ is an odd function of the
components of the wave vector. It is not equal to zero if the structure
lacks inversion symmetry (which is very typical for nanostructures). This is
the case when either the crystal unit cell lacks inversion symmetry (bulk
inversion asymmetry, BIA \cite{Dresselhaus}), or the QW is asymmetric
(structure inversion asymmetry, SIA \cite{Rashba}). The components of ${\bf h%
}({\bf k})$ may be, or may not be, associated with certain Cartesian axes in
the real space. In the two-dimensional case, ${\bf h}({\bf k})$ is dominated
by linear in ${\bf k}$ terms \cite{Rashba,Aronov,DyakKach}:

\begin{equation}
h_{\eta }=A_{\eta \zeta }k_{\zeta }  \label{4}
\end{equation}

where the matrix $A$ is defined by the structure symmetry.

The problem we are going to solve is finding the fine structure of the
ground state of the two-particle Hamiltonian:

\begin{equation}
\hat{H}=\hat{H}_{0}+\hat{H}_{SO}  \label{5}
\end{equation}

where

\begin{equation}
\hat{H}_{0}=\frac{\hbar ^{2}}{2m}\hat{k}_{1}^{2}+\frac{\hbar ^{2}}{2m}\hat{k}%
_{2}^{2}+V_{1}({\bf r}_{1})+V_{2}({\bf r}_{2})+U_{12}(\left| {\bf r}_{1}-%
{\bf r}_{2}\right| )  \label{6}
\end{equation}

\begin{equation}
\hat{H}_{SO}={\bf h}({\bf k}_{1})\cdot {\bf j}_{1}{\bf +h}({\bf k}_{2})\cdot 
{\bf j}_{2}  \label{7}
\end{equation}

and $U_{12}(\left| {\bf r}_{1}-{\bf r}_{2}\right| )$ is the operator of the
Coulomb interaction between the two particles. Before tackling the effects
of spin-orbit interaction in the form of Eq.(7) on the exchange interaction,
we should reconsider the ground-state structure of the Hamiltonian $H_{0}$.
It is indeed well-known for electrons whose one-particle wave functions are $%
\Psi ({\bf r})\zeta _{\mu }$, where the spinor $\zeta _{\mu }$ is not a
function of coordinates. To the contrary, the Bloch amplitude $u_{\nu }$
does depend on coordinates, and, moreover, it may contain spinors with both $%
\mu =+1/2$ and $\mu =-1/2$. The exciton (an electron-hole pair) in a QD is a
good example demonstrating that the exchange interaction of charge carriers
may have a very different symmetry as compared to that of free electrons.
The QD exciton fine structure \cite{IvchPik} consists of two doublets, being
thus quite different from the fine structure of a pair of vacuum electrons,
i.e. the well-known singlet-triplet structure associated with the Heisenberg
exchange.

In order to analyze the fine structure of $H_{0}$ for two particles
belonging to the same subband, we first note that their behavior should be
identical to that of bare electrons in all aspects but the Coulomb
interaction. Indeed, although Bloch amplitudes are functions of coordinates
within the unit cell, the one-particle operators of the kinetic energy and
of the potential energy in Eq.(6) in the effective-mass approximation act
upon envelope function, not Bloch amplitudes. Therefore, with respect to
these one-particle operators, the Bloch amplitudes are just equivalent to
spinors. To the contrary, calculating the Coulomb energy assumes taking
integrals over the unit sell also. It is due this fact that the symmetry of
Bloch amplitudes of holes and electrons has an impact on the fine structure
in the exciton \cite{IvchPik,Goupalov}.

The fermionic wave functions of the two charge carriers can be written in
the following form, similar to that of bare electrons:

\begin{eqnarray}
\Psi _{0}({\bf r}_{1},{\bf r}_{2}) &=&\left[ \Phi _{0}({\bf r}_{1},{\bf r}%
_{2})+\Phi _{0}({\bf r}_{2},{\bf r}_{1})\right] \xi _{0}  \nonumber \\
\Psi _{1M}({\bf r}_{1},{\bf r}_{2}) &=&\left[ \Phi _{1}({\bf r}_{1},{\bf r}%
_{2})-\Phi _{1}({\bf r}_{2},{\bf r}_{1})\right] \xi _{1M}  \label{8}
\end{eqnarray}

where $\Phi _{0}({\bf r}_{1},{\bf r}_{2})$ and $\Phi _{1}({\bf r}_{1},{\bf r}%
_{2})$ are two-particle envelope functions defined so that each particle is
most likely to be found near its ''home'' center, while $\Phi _{0}({\bf r}%
_{2},{\bf r}_{1})$ and $\Phi _{1}({\bf r}_{2},{\bf r}_{1})$ correspond to
interchanged particle positions. To determine the structure of respective
energy levels, we should recall a property of the Bloch amplitudes $u_{\nu }$%
, which results from their symmetry with respect to time reversal, and is an
equivalent formulation of the Kramers theorem. As it follows from the
Kramers theorem, the states symmetric with respect to time reversal remain
degenerated unless magnetic field is applied. Mathematically, this means
that matrix elements of any function of coordinates (not containing
derivatives or spin operators) between $u_{+1/2}({\bf r})$ and $u_{-1/2}(%
{\bf r})$ are zero, while diagonal matrix elements are equal to each other:

\begin{eqnarray}
\int\limits_{\Omega }u_{+1/2}({\bf r})u_{-1/2}^{*}({\bf r})f({\bf r})d^{3}r
&=&\int\limits_{\Omega }u_{-1/2}({\bf r})u_{+1/2}^{*}({\bf r})f({\bf r}%
)d^{3}r=0  \nonumber \\
\int\limits_{\Omega }u_{+1/2}({\bf r})u_{+1/2}^{*}({\bf r})f({\bf r})d^{3}r
&=&\int\limits_{\Omega }u_{-1/2}({\bf r})u_{-1/2}^{*}({\bf r})f({\bf r}%
)d^{3}r  \label{9}
\end{eqnarray}

where the integrals are taken over the unit cell.

Using Eq.(9), one can easily find that

\begin{eqnarray}
\left\langle \Psi _{0}({\bf r}_{1},{\bf r}_{2})\right| U_{12}(\left| {\bf r}%
_{1}-{\bf r}_{2}\right| )\left| \Psi _{1M}({\bf r}_{1},{\bf r}%
_{2})\right\rangle &=&0  \nonumber \\
\left\langle \Psi _{0}({\bf r}_{1},{\bf r}_{2})\right| U_{12}(\left| {\bf r}%
_{1}-{\bf r}_{2}\right| )\left| \Psi _{0}({\bf r}_{1},{\bf r}%
_{2})\right\rangle &\neq &  \label{10} \\
\left\langle \Psi _{1M}({\bf r}_{1},{\bf r}_{2})\right| U_{12}(\left| {\bf r}%
_{1}-{\bf r}_{2}\right| )\left| \Psi _{1M^{\prime }}({\bf r}_{1},{\bf r}%
_{2})\right\rangle &=&const\cdot \delta _{MM^{\prime }}  \nonumber
\end{eqnarray}

Thus, the Coulomb interaction retains the singlet-triplet structure of the
ground state of two identical charge carriers. Exactly like in the case of
two bare electrons, two-particle states with the same total pseudospin $I$
are degenerated. Consequently, the Hamiltonian of the exchange interaction
in terms of pseudospin operators takes the Heisenberg form:

\begin{equation}
\hat{H}_{S}=-2\Delta ({\bf j}_{1}{\bf \cdot j}_{2}+1{\bf )}  \label{11}
\end{equation}

where $\Delta $ is a constant to be determined for each specific case.

Now we can consider the effect of the spin-orbit terms given by the Eq.(7)
on the exchange interaction. In the following, we will choose the axis X
along the straight line connecting the localization centers (QDs). To handle
the spin-orbit terms, we make use of a unitary transformation proposed by
Levitov and Rashba \cite{LevRash} who used it to eliminate spin-orbit terms
in the one-dimensional case. The matrix $T$ defined as

\begin{equation}
T=\exp \left[ i\frac{2m}{\hbar ^{2}}\sum\limits_{\alpha }A_{\alpha x}(\hat{%
\jmath}_{1\alpha }x_{1}+\hat{\jmath}_{2\alpha }x_{2})\right]  \label{12}
\end{equation}

transforms the Hamiltonian Eq.(5) into the form:

\begin{equation}
T\hat{H}T^{-1}=\hat{H}^{\prime }=\hat{H}_{0}^{\prime }+\hat{H}_{SO}^{\prime }
\label{13}
\end{equation}

where

\begin{equation}
\hat{H}_{0}^{\prime }=\frac{\hbar ^{2}}{2m}\hat{k}_{1}^{\prime 2}+\frac{%
\hbar ^{2}}{2m}\hat{k}_{2}^{\prime 2}+V_{1}({\bf r}_{1}^{\prime })+V_{2}(%
{\bf r}_{2}^{\prime })+U_{12}(\left| {\bf r}_{1}^{\prime }-{\bf r}%
_{2}^{\prime }\right| )-\sum\limits_{\alpha }\frac{m(A_{\alpha x})^{2}}{%
\hbar ^{2}}  \label{14}
\end{equation}

and

\begin{equation}
\hat{H}_{SO}^{\prime }=\sum\limits_{\alpha }A_{\alpha y}(k_{1y}\hat{\jmath}%
_{1\alpha }^{\prime }+k_{2y}\hat{\jmath}_{2\alpha }^{\prime })  \label{15}
\end{equation}

where ${\bf j}_{1}^{\prime }=T{\bf j}_{1}T^{-1}$, ${\bf j}_{2}^{\prime }=T%
{\bf j}_{2}T^{-1}$.

The Hamiltonian $\hat{H}^{\prime }$ does not contain spin-orbit terms and
therefore results in the exchange interaction in the form of Eq.(11):

\begin{equation}
\hat{H}_{S}^{\prime }=-2\Delta ({\bf j}_{1}^{\prime }{\bf \cdot j}%
_{2}^{\prime }+1{\bf )}  \label{16}
\end{equation}

Due to the axial symmetry of the system, the matrix elements of $k_{1y}$ and 
$k_{2y}$, calculated on the ground-state eigenfunctions of $\hat{H}%
_{0}^{\prime }$ (they can be obtained from Eq.(8) by the transformation with
the matrix $T$, which does not affect their dependence on $y$), are exactly
equal to zero. The same is true for all the odd powers of $k_{1y}$ and $%
k_{2y}$. Therefore, $\hat{H}_{SO}^{\prime }$ does not contribute into the
exchange interaction.

Finally, to obtain the exchange Hamiltonian in the non-transformed basis,
one should substitute the expressions for ${\bf j}_{1}^{\prime }$ and ${\bf j%
}_{2}^{\prime }$ into Eq.(16). Since the transformation $T$ is a rotation
through the angle $\frac{2m}{\hbar ^{2}}\sqrt{\sum\limits_{\nu }A_{\nu
x}A_{\nu x}}x$ around the vector $A_{\nu x}$ in the pseudospin space, $\hat{H%
}_{S}$ in the non-transformed basis is not unambiguously defined: it depends
on the coordinates $x_{1}$ and $x_{2}$ at which we take the spin operators $%
{\bf j}_{1}$ and ${\bf j}_{2}$. A natural choice is to define them at the
centers of corresponding QDs; for instance, this definition allows to write
the Zeeman interaction in the usual form, $\hat{H}_{Z}=\mu _{B}g_{\alpha
\beta }j_{\alpha }B_{\beta }$, where $B$ is the magnetic field, $\mu _{B}$
is the Bohr magneton, and $g_{\alpha \beta }$ is a symmetric tensor g-factor%
\cite{AbrBlin} whose principal directions do not depend on the envelope wave
function of the localized particle. This way, we come to the expression for $%
\hat{H}_{S}$ obtained in Ref.\cite{Moriarty}:

\begin{equation}
\hat{H}_{S}=-2\Delta (1+{\bf j}_{1}{\bf \cdot j}_{2}\cos \gamma +({\bf %
d\cdot j_{1})}({\bf d\cdot j_{2})(1-\cos \gamma )+d\cdot }\left[ {\bf j}%
_{1}\times {\bf j}_{2}\right] \sin \gamma )  \label{17}
\end{equation}

where $\gamma =\frac{2m}{\hbar ^{2}}\sqrt{\sum\limits_{\nu }A_{\nu x}A_{\nu
x}}R_{12}$, and ${\bf d}$ is a unit vector in the pseudospin space, defined
so that $d_{\nu }=A_{\nu x}/\sqrt{\sum\limits_{\nu }A_{\nu x}A_{\nu x}}$.
The first anisotropic term has the form of pseudodipole interaction \cite
{AbrBlin}, and the second one, of the Dzyaloshinskii-Moriya interaction \cite
{Dzialoshinskii}. At small $\gamma $, the Dzyaloshinskii-Moriya interaction
dominates.

The Eq.(17) demonstrates a remarkable universality of the exchange
interaction in two-dimensional semiconductor nanostructures: this form of
the Hamiltonian holds for both electrons and holes, for any type of
centrosymmetric localizing potentials. The Eq.(17) is valid for identical as
well as for different QDs. Moreover, the angle $\gamma $ characterizing the
relative strength of the anisotropic exchange depends only on the distance
between the QDs and the orientation of the pair of QDs with respect to the
crystal axes. It is not sensitive to binding energies of the charge carriers
in the QDs and to the value of the isotropic exchange constant $\Delta $.

The value of $\gamma $ can be now easily calculated for those structures
where the components of the matrix $A$ are known.

In [100] oriented GaAs quantum wells the dominating BIA terms are \cite
{DyakKach} $A_{yy}=-A_{xx}=\frac{\alpha \hbar ^{3}}{m\sqrt{2mE_{g}}}%
\left\langle k_{z}^{2}\right\rangle $, $A_{xy}=A_{yx}=0$ where $\alpha
\approx 0.65$ \cite{Juss} (here coordinates $x$ and $y$ are taken along the
cubic crystal axes). This gives $\gamma =\frac{2\alpha \hbar }{\sqrt{2mE_{g}}%
}\left\langle k_{z}^{2}\right\rangle R_{12}$. For example, in a 5nm-wide QW
considered in Ref.\cite{BurLoss}, $\gamma =(3\cdot 10^{5}cm^{-1})R_{12}$,
which gives for $R=24nm$ $\gamma =0.72$, i.e. more than an order of
magnitude greater than the value of 0.02 obtained in Ref.\cite{BurLoss} at
the same values of parameters by the Heitler-London method.

For Rashba terms, $A_{yy}=A_{xx}=0$, $A_{xy}=-A_{yx}=a$. In a single-side
modulation-doped n-type Si/SiGe quantum well, the constant $a$ of 1.1$\cdot $%
10$^{-12}$ eV$\cdot $cm was measured \cite{Will}. This gives $\gamma =\frac{%
2ma}{\hbar ^{2}}R_{12}\approx (6.7\cdot 10^{2}cm^{-1})R_{12}$.

Bulk inversion asymmetry terms for holes in zinc-blende semiconductors
include both cubic and linear in $k$ terms\cite{PikTit}. The cubic term $%
H_{3V}$ has the same symmetry as the Dresselhaus term for electrons, with
the constant $\frac{\alpha _{V}\hbar ^{3}}{m\sqrt{2mE_{g}}}$, where $m$ is
the conduction-band electron mass, and $\alpha _{V}\approx 0.1$ for GaAs.
The linear term is given by the expression:

\begin{equation}
H_{1V}=\frac{4}{\sqrt{3}}\varkappa \left( {\bf k\cdot \Omega }\right)
\label{18}
\end{equation}

where $\Omega _{z}=\hat{J}_{z}\left( \hat{J}_{x}^{2}-\hat{J}_{y}^{2}\right)
+\left( \hat{J}_{x}^{2}-\hat{J}_{y}^{2}\right) \hat{J}_{z}$ (other
components of ${\bf \Omega }$ are obtained by cyclic interchange of
indices), and $\varkappa \approx 10^{-10}eV\cdot cm$. Taking matrix elements
of $H_{1V}$ and $H_{3V}$ within pairs of the states with $J_{z}=\pm 1/2$
(light holes) and $J_{z}=\pm 3/2$ (heavy holes), and going to the pseudospin
notation, we obtain for a [100] QW:

\begin{eqnarray}
{\bf h}_{l}({\bf k}) &=&-\left( 2\frac{\alpha _{V}\hbar ^{3}}{m\sqrt{2mE_{g}}%
}\left\langle k_{z}^{2}\right\rangle -2\sqrt{3}\varkappa \right) \left( k_{x}%
\hat{\jmath}_{x}-k_{y}\hat{\jmath}_{y}\right)  \nonumber \\
{\bf h}_{h}({\bf k}) &=&-4\sqrt{3}\varkappa \left( k_{x}\hat{\jmath}%
_{x}-k_{y}\hat{\jmath}_{y}\right)  \label{19}
\end{eqnarray}

where ${\bf h}_{l}({\bf k})$ and ${\bf h}_{h}({\bf k})$ are spin-orbit
fields (see Eq.(3)) for light and heavy holes respectively. Consequently, $%
\gamma _{l}=\left( \frac{4\alpha _{V}\hbar m_{l}}{m\sqrt{2mE_{g}}}%
\left\langle k_{z}^{2}\right\rangle -\frac{4\sqrt{3}m_{l}}{\hbar ^{2}}%
\varkappa \right) R_{12}$ for light holes, and $\gamma _{h}=\frac{8\sqrt{3}%
m_{h}}{\hbar ^{2}}\varkappa R_{12}$ for heavy holes, where $m_{l}$ and $%
m_{h} $ are effective masses of the light and heavy hole respectively,
corresponding to their motion along the QW plane. For example, in a
10nm-wide GaAs QW, $\gamma _{l}\approx \gamma _{h}\approx (2\cdot
10^{5}cm^{-1})R_{12}$.

The symmetry of exchange interactions has been discussed in relation to
feasibility of quantum computation with spins of localized electrons in
semiconductor nanostructures \cite{Moriarty,Bonesteel,BurLoss}. A necessary
(but not sufficient; see Ref.\cite{Dyakonov}) condition for practical
quantum computing to become possible is that the error probability per
quantum gate be less than a certain value (of the order of 10$^{-5}$) \cite
{Preskill}. As shown in Ref.\cite{BurLoss}, there exists a way of performing
exchange-mediated quantum gates that allows to avoid errors caused by the
anisotropy, provided $\gamma $ remains constant when the isotropic exchange
constant is changed. The above consideration shows that this is indeed the
case as long as spin-orbit constants do not change. They may change,
however, because application of electric fields to the structure can either
alter $\left\langle k_{z}^{2}\right\rangle $ or bring about SIA (Rashba)
terms. Since typical values of $\gamma $ are in reality much greater than
estimated in Ref.\cite{BurLoss}, the uncontrollable effect of anisotropy on
exchange-mediated quantum gates can not be so easily discarded.

In conclusion, the exchange interaction of charge carriers (electrons or
holes) localized in two-dimensional semiconductor structures is shown to be
described by an universal Hamiltonian in terms of carriers' pseudospins. It
has the Heisenberg form unless spin-orbit terms, linear in the carrier wave
vector, are present in the total Hamiltonian of the system. In this latter
case, anisotropic contributions having both Dzyaloshinskii-Moriya and
pseudodipole form arise. The ''rotation angle'' $\gamma $, characterizing
the relative strength of the anisotropic exchange, linearly depends on the
distance between the localization centers and does not depend on binding
energies of the carriers.

The author is grateful to I.A.Merkulov, E.L.Ivchenko, L.P.Gor'kov, and
P.G.Krotkov for stimulating discussions. Partial support of INTAS, RFBR, and
UK EPSRC is acknowledged.

\end{document}